\pdfoutput=1
\documentclass[%
prb,%
letterpaper,%
twocolumn,%
superscriptaddress,%
showpacs,%showkeys,%
amsmath,%
amssymb%
]{revtex4}

\newcommand{\ket}[1]{\ensuremath{\left\vert #1 \right\rangle}}

\newcommand{\average}[1]{\ensuremath{\left\langle #1 \right\rangle}}
\newcommand{\matrixelement}[3]{\ensuremath{\left\langle #1 \vert #2 \vert #3 \right\rangle}}
\newcommand{\genpyro}[3]{\ensuremath{\text{#1}_{2}\text{#2}_{2}\text{O}_{6}\text{#3}}}
\newcommand{\pyro}[2]{\ensuremath{\text{#1}_{2}\text{#2}_{2}\text{O}_{7}}}
\newcommand{\bpyro}[2]{\ensuremath{\text{#1}\text{#2}_{2}\text{O}_{6}}}
\newcommand{\temperature}[1]{\ensuremath{\mathit{T}_{#1}}}

\newcommand{\kelvin}{\ensuremath{\text{K}}}
\newcommand{\percent}{\ensuremath{\%}}

\usepackage{hyperref}
\usepackage{graphicx}% Include figure files
\begin{document}

%% REVTeX style
%\begin{CJK*}{UTF8}{}

\title{Partial order in a frustrated Potts model}

%% JPSJ style
% \author{Ryo \textsc{Igarashi}$^{1,2}$\thanks{E-mail: igarashi.ryo@jaea.go.jp}, and Masao \textsc{Ogata}$^{3}$}
% \inst{%
%   $^{1}$CCSE, Japan Atomic Energy Agency, Higashi-Ueno, Tokyo 110-0015\\
%   $^{2}$CREST(JST), Honcho, Kawaguchi, Saitama 443-0012\\
%   $^{3}$Department of Physics, University of Tokyo, Hongo, Tokyo 133-0033}

%% REVTeX style
%\CJKfamily{min}
%\author{Ryo Igarashi (五十嵐 亮)}
\author{Ryo Igarashi}
\email{igarashi.ryo@jaea.go.jp}
\affiliation{CCSE, Japan Atomic Energy Agency, Higashi-Ueno, Taito-ku, Tokyo 110-0015, Japan}
\affiliation{CREST(JST), Honcho, Kawaguchi, Saitama 443-0012, Japan}
%\author{Masao Ogata (小形 正男)}
\author{Masao Ogata}
\affiliation{Department of Physics, University of Tokyo, Hongo, Bunkyo-ku, Tokyo 133-0033, Japan}

%% REVTeX style
\date{\today}

%% JPSJ style
%\abst{%
%% REVTeX style
\begin{abstract}%
  We investigate a 4-state ferromagnetic Potts model
  with a special type of geometrical frustration
  on a three dimensional diamond lattice
  by means of Wang-Landau Monte Carlo simulation
  motivated by a peculiar structural phase transition
  found in $\beta$-pyrochlore oxide $\bpyro{K}{Os}$.
  We find that this model undergoes unconventional first-order phase transition;
  half of the spins in the system order in a two dimensional
  hexagonal-sheet-like structure, while the remaining half stay disordered.
  The ordered sheets and the disordered sheets stack one after another.
  We obtain a fairly large residual entropy at $\temperature{} = 0$ which originates
  from the disordered sheets.
%% JPSJ style
%}
%% REVTeX style
\end{abstract}

%% JPSJ style
%\kword{Potts model, Monte Carlo method, Wang-Landau algorithm, frustration}
%% REVTeX style
%\pacs{75.10.Hk, 05.50.+q, 05.10.Ln, 02.70.Uu, 64.60.De, 64.60.Ej, 63.70.+h, 63.20.kg}
\pacs{75.10.Hk, 05.50.+q, 05.10.Ln, 02.70.Uu}
%\keywords{Potts model, Monte Carlo method, Wang-Landau algorithm, frustration}

%% JPSJ style
%\begin{document}

\maketitle

%% REVTeX style
%\end{CJK*}

\section{Introduction}
\label{cha:introduction}

Generally speaking, frustrated systems have some constraints
that forbid simultaneous minimization of all the interaction energies.
Therefore, frustration usually suppresses phase transitions
to long-range orders, and,
as a result, leads to very rich low temperature phases.
Moreover, a frustrated system may become a spin-liquid phase
or sometimes exhibit  phase transition to a partially ordered state.
In this paper, we study a 4-state ferromagnetic Potts model
with a special type of geometrical frustration on a three-dimensional
diamond lattice.
This model is proposed as a simplified model Hamiltonian
for a peculiar phase transition found in $\beta$-pyrochlore oxide $\bpyro{K}{Os}$.
Although the obtained phase transition in this model
does not explain that in $\bpyro{K}{Os}$,
we find that it has several interesting properties as a model
with geometrical frustration.

Let us briefly summarize experimental results of $\beta$-pyrochlore superconductors,
$\bpyro{A}{Os}$ (A is one of K, Rb or Cs)~\cite{Hiroi2005Second,Yonezawa2004New,Yonezawa2004Newa,Yonezawa2004Superconductivity},
from which the particular model
studied in this paper is derived.
We mainly focus on the anomalous phase transition
other than superconducting transition in $\bpyro{K}{Os}$.
The pyrochlore structure in general is of type $\genpyro{A'}{B'}{O'}$
with cubic space-group $Fd\overline{3}m$\cite{Yamaura2006Crystal,Schuck2006Crystal}.
The four crystallographically inequivalent atoms $\text{A'}$, $\text{B'}$, $\text{O}$, and $\text{O'}$
occupy the $16c$, $16d$, $48f$, and $8b$ sites, respectively,
in the fcc unit cell.
The crystal structure of the $\beta$-pyrochlore
is derived from the general pyrochlore structure
by removing $\text{A'}$ atoms,
by replacing $\text{O'}$ (oxygen) atoms
with alkali metal atoms,
and by filling the $\text{B'}$ position with osmium atoms.
The superconducting transition temperature, $\temperature{c}$,
of this $\beta$-pyrochlore oxide family is rather high~\cite{Yonezawa2004Newa,Yonezawa2004New,Yonezawa2004Superconductivity,Bruhwiler2004Superconductivity,Kazakov2004Synthesis}
compared with the previously discovered pyrochlore oxide
superconductor, $\pyro{Cd}{Re}$~\cite{Hanawa2001Superconductivity}.

Several first-principle
density functional calculations
were performed for
the various $\beta$-pyrochlore compounds~\cite{Saniz2004Electronic,Kunevs2004Correlation}.
The electric structure
is not affected
by changing the alkali atoms
which are almost univalent.
The osmium atoms and the oxygen atoms form
OsO$_{6}$ octahedra and OsO$_{6}$ network acts
as cages for the alkali atoms.
The cage size
is also insensitive to the alkali atom content~\cite{Yonezawa2004New,Yamaura2006Crystal}.
Therefore, the difference of $\temperature{c}$
among the $\beta$-pyrochlore family
comes from the size mismatch
between the cages and the alkali atoms.
The heaviest anharmonic oscillation called ``rattling''
is observed for the smallest alkali atom contents,
$\text{K}$\cite{Yamaura2006Crystal,Hiroi2005Specific,Kunevs2004Correlation}.
The spatial asymmetry of the electron density of $\text{K}$ atoms
is also observed~\cite{Yamaura2006Crystal,Galati2007structure},
which favors the nearest neighbor $\text{K}$ atom direction.

Aside from the superconductivity,
the potassium based $\beta$-pyrochlore $\bpyro{K}{Os}$
undergoes a novel phase transition
near $7.6\kelvin$~\cite{Hiroi2005Second,Hiroi2005Erratum,Bruhwiler2006Mass}.
This is a first-order phase transition
where both specific heat and resistivity show hysteresis~\cite{Hiroi2007Extremely,Hiroi2007second}.
Surprisingly, its transition temperature
is independent of the superconducting $\temperature{c}$.
Therefore, this transition is not an electric origin
and is assumed to be of structural one
which relates to the ``rattling'' of the $\text{K}$ atoms.
However, no evidence for lattice distortion has been found
in X-ray,
NMR~\cite{Yoshida2007NMR,Yoshida2007NMRa},
high pressure transport measurements~\cite{Akrap2007Manifestations},
scanning tunneling spectroscopy~\cite{Dubois2008Scanning},
the photo-emission spectroscopy~\cite{Shimojima2007Interplay}
or Raman spectroscopy~\cite{Hasegawa2007Raman,Hasegawa2008Raman,Schoenes2008Phonons}
measurements.
Thus, in spite of the very sharp transition,
the nature of the phase transition remains a big mystery.

\section{Model}
\label{sec:model}

Having the peculiar first-order phase transition of $\bpyro{K}{Os}$ described above in mind,
we study a simple classical model
with special kind of geometrical frustration,
which was proposed by
Kun{\v e}s \textit{et al.}
using the density functional calculation~\cite{Kunevs2006Effective,Kunevs2006Frustration,Kunevs2006KOs_2O_6,Kunevs2004Correlation}.
They investigated that
the on-site potential of the $\text{K}$ atoms
along with $\text{K}$--$\text{K}$ bond direction is heavily unharmonic
and very flat near the symmetry center,
which is formed by osmium and oxygen surrounding the $\text{K}$ ion is over-sized.

\begin{figure}
  \begin{center}
    \includegraphics[width=\columnwidth,clip,keepaspectratio]{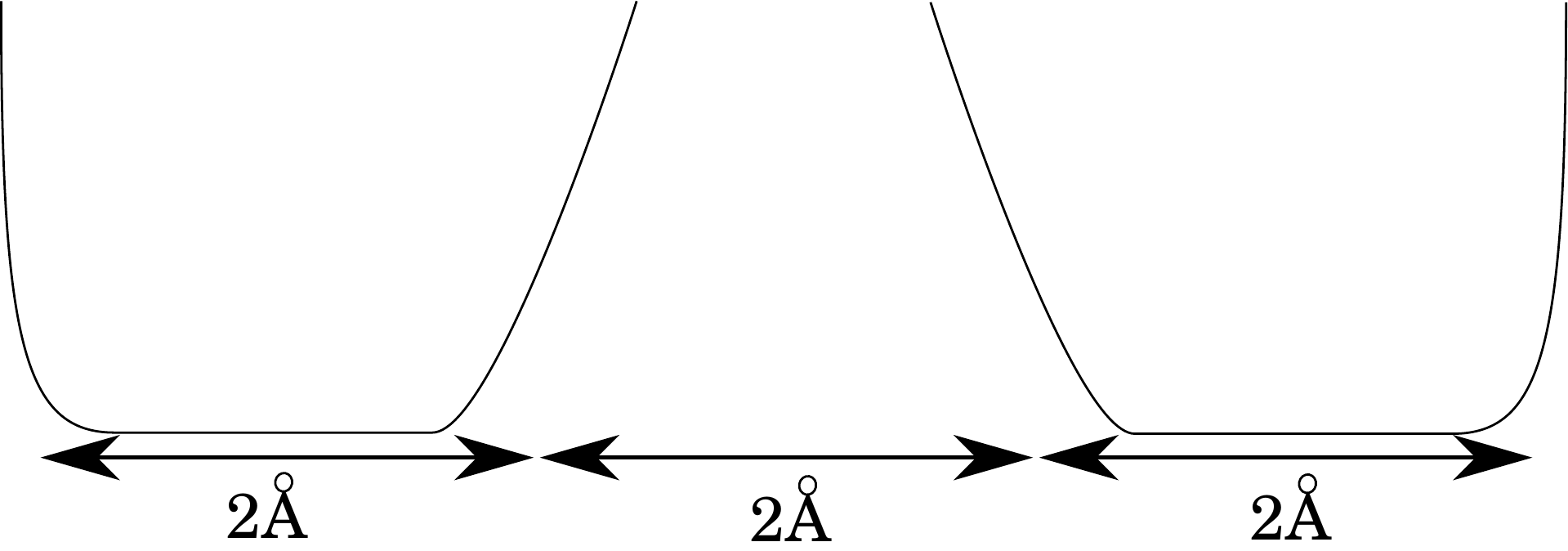}
  \end{center}
  \caption{
    An approximate length-scale image of the potential wells
    for $\text{K}$ atoms.
    The bond center is the inversion symmetry center.
  }
  \label{fig:approximate}
\end{figure}

The cages formed by $\text{OsO}_{6}$ octahedra
in the $\beta$-pyrochlore structure
have 4 holes towards nearest-neighbor $\text{K}$ atoms
because the $16c$ sites are empty.
Therefore these four holes in the cages are
not only the origin of anisotropic potential of a $\text{K}$ atom but
also the origin of rather strong Coulomb interaction between $\text{K}$ atoms.
The $\text{K}$ atoms in the $\beta$-pyrochlore $\bpyro{K}{Os}$ form
the diamond lattice structure.
Thus the coupling between the nearest-neighbor $\text{K}$ atoms is essentially of 
Coulomb origin and repulsive.
Actually, as shown in the approximate length scale of
the cage and the flat potential region of the $\text{K}$ atoms
in Fig.~\ref{fig:approximate},
the distance between cages is very small.

For discussing $\text{K}$ atoms quantum mechanically,
it is useful to use the states $\ket{\alpha}$ ($\alpha = 1, 2, 3, 4$)
as the basis
which represents the $sp^3$-like wavefunctions made out of $1s$ and $2p$ states,
each located towards the holes of the cages.
The cages
will play the roll of metallic screening
and thus we assume that the further long-range interaction can be neglected. 
\begin{figure}
  \centering
  \includegraphics[width=0.49\columnwidth,clip,keepaspectratio]{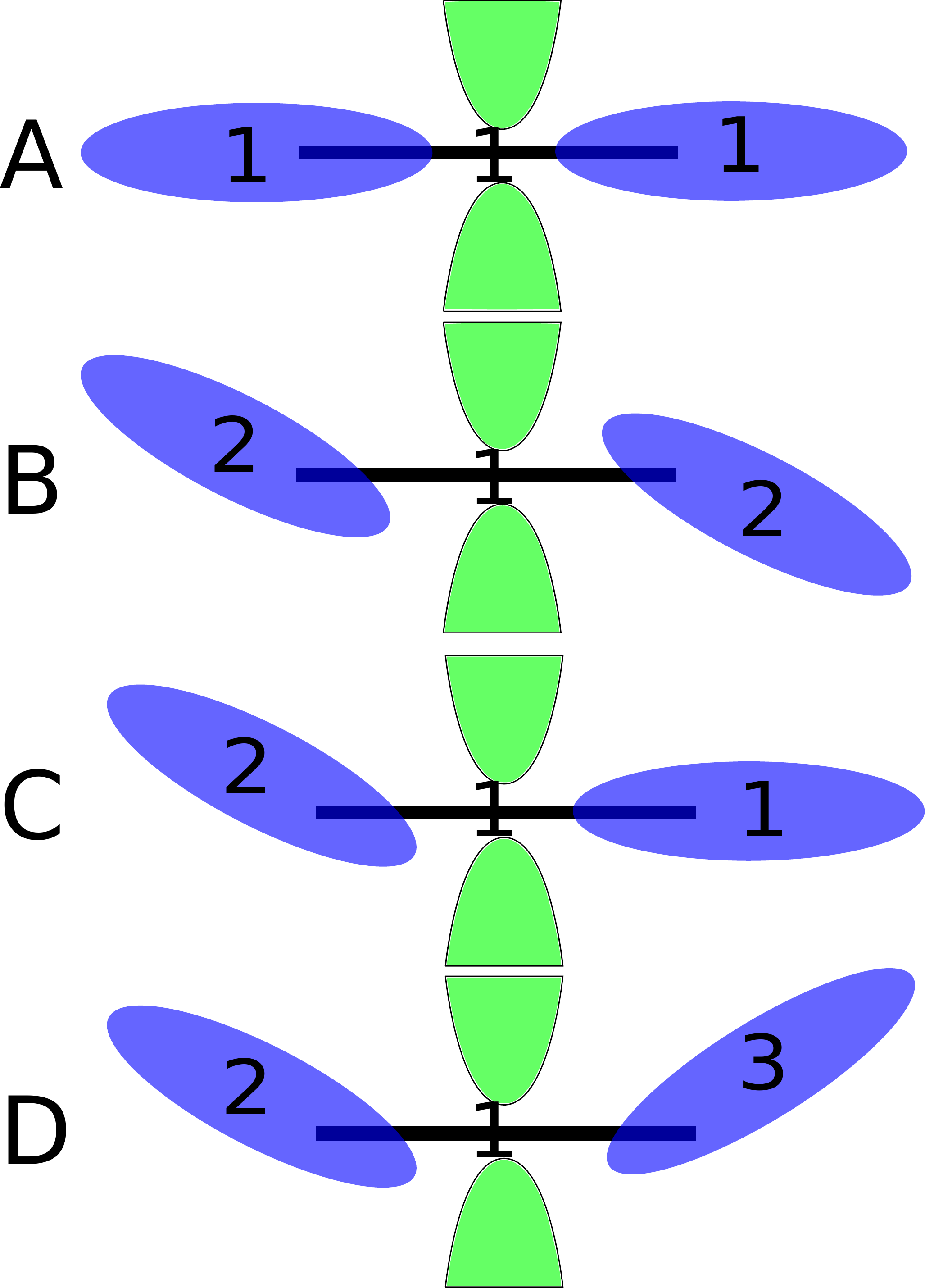}
  \caption{
    (Color Online) Some examples of configurations of ``bond number'', $R(ij)$,
    and Potts variable $\alpha$ at both ends.
    Each oval means that one of the four wavefunctions,
    $\alpha = 1, 2, 3, 4$, extending in some direction.
    The green structures represent that the $\text{Os}$--$\text{O}$ cage have holes towards 
    the next-nearest $\text{K}$ atoms.
    Configuration A: two wavefunctions and the bond are all parallel.
    This gives an energy of $- 2J_{1} + 2J_{2}$
    and is energetically unfavored since $J_{2} > J_{1}$.
    Configuration B: two wavefunctions are parallel but the bond direction is different case.
    This gives an energy of $-2J_{1}$ and is energetically favored.
    Configuration C: one of the spins and the bond are parallel but the other spins is not. This gives an energy of
    $J_{3}$.
    Configuration D gives an energy of 0.}
  \label{fig:fourspinconfig}
\end{figure}
The flat potential and the holes in cages lessen
the energy difference between the singlet and triplet states
and increases the energy difference to the higher energy states.
Therefore we focus on the lowest four states and neglect 
the energy difference between the lowest singlet and triplet states.
In this case we can assume the four-fold  degenerate basis states, 
$\ket{\alpha}$ ($\alpha = 1, 2, 3, 4$), for each $\text{K}$ atom in every cage.
These states can be represented in the $sp^{3}$-like structure
which point to the nearest-neighbor $\text{K}$ atoms.  
We assign these states as the four states of the classical Potts model.
Note that we have one-to-one correspondence between
the Potts variable 1 to 4 and the direction
in the diamond lattice.

We concentrate on the inter-site Coulomb couplings
represented in the lowest 4 states.
Although
there are
$4^4$ matrix elements, $\matrixelement{\alpha \beta}{W}{\alpha' \beta'}$,
the largest contribution comes from
the diagonal matrix elements,
$\matrixelement{\alpha \beta}{W}{\alpha \beta}$~\cite{Kunevs2006KOs_2O_6,Kunevs2006Effective,Kunevs2006Frustration}.
These  matrix elements are estimated to be 
\begin{align}
  \label{eq:27}
  \matrixelement{\alpha \beta}{W_{R}}{\alpha \beta}
  = -2J_{1} \delta_{\alpha \beta}
  + 2J_{2} \delta_{\alpha \text{R}} \delta_{\beta \text{R}}
  - J_{3} (\delta_{\alpha \text{R}} + \delta_{\beta \text{R}})
  \\
  ( J_{1} = 162\kelvin,
  J_{2} = 371\kelvin,
  J_{3} = 301\kelvin ) \notag
\end{align}
where
$\alpha$ and $\beta$ denotes the direction of the bond,
$\delta_{\alpha \beta}$ is the Kronecker delta
and $R$ is a so-called ``bond number'',
which represents the 4 type of bond directions in the diamond lattice structure
and takes the value of 1 to 4.
For example, $R=1$ means that the direction of the bond is
$\average{111}$, and similarly
the basis state with $\alpha = 1$
represents one of the four wavefunctions extending
in the $\average{111}$ direction.
4 types of typical configurations are shown in Fig.~\ref{fig:fourspinconfig}.
The first term in the right-hand-side of eq.~\eqref{eq:27}
represents the energy gain of $2J_{1}$ when $\text{K}$ atoms
on the both ends of a bond are in the same direction
$\alpha = \beta$.
For example, the configuration A and B in Fig.~\ref{fig:fourspinconfig}
corresponds to this case.
The second term in the right-hand-side of eq.~\eqref{eq:27} 
means that there is an energy loss of $2J_{2}$ when the bond 
direction ($R$) and the directions of the states $\ket{\alpha}$ 
and $\ket{\beta}$ at the both ends of the bonds are all in a straight line.  
The configuration A in Fig.~\ref{fig:fourspinconfig} also corresponds to this case.
The third term in the right-hand-side of eq.~\eqref{eq:27}
represents the energy gain of $J_{3}$ when the wavefunction of one side
is parallel to the bond direction. This corresponds to the configuration C
in Fig.~\ref{fig:fourspinconfig}.
As a result, configuration A in Fig.~\ref{fig:fourspinconfig} has an energy,
$-2J_{1} + 2J_{2}$, configuration B, $-2J_{1}$, configuration C, $-J_{3}$
and configuration D, zero.
We can neglect the $J_{3}$ terms because these terms corresponds to the constant value
after summation of all bonds.

The off-diagonal terms,
such as 
$\matrixelement{\alpha \beta}{W}{\alpha \gamma}$, is of order
smaller than $J_{1}$, $J_{2}$ and $J_{3}$.
Therefore we neglect the off-diagonalized terms in the following.  
Furthermore, we assume that the effects of the
excited states higher than the
lowest four states are negligible.
Finally, the effective Hamiltonian
for the inter-site couplings of $\text{K}$ atoms becomes
\begin{align}
  \label{eq:28}
  H = \sum_{\average{ij}}
  \left(
    -J_{1} \delta_{\alpha \beta}^{ij} +
    J_{2} \delta_{\alpha R(ij)}^{i} \delta_{\beta R(ij)}^{j}
  \right),
\end{align}
where $R(ij)$ is also a so-called ``bond number'', 1--4,
between 
$i$-site and $j$-site, which takes the same value of $R$
described above.
$J_{1}$ term shows energy gain of parallel oscillation of $\text{K}$ ion
and $J_{2}$ term shows energy loss of oscillation in line.
The bare values
of $J_{1}$ and $J_{2}$ are $162\kelvin$ and 
$371\kelvin$, respectively~\cite{Kunevs2006Frustration}.
The sum is taken over the $\text{K}$--$\text{K}$ bond network
forming the diamond lattice structure.

\begin{figure}
  \centering
  \includegraphics[width=0.5\columnwidth,clip,keepaspectratio]{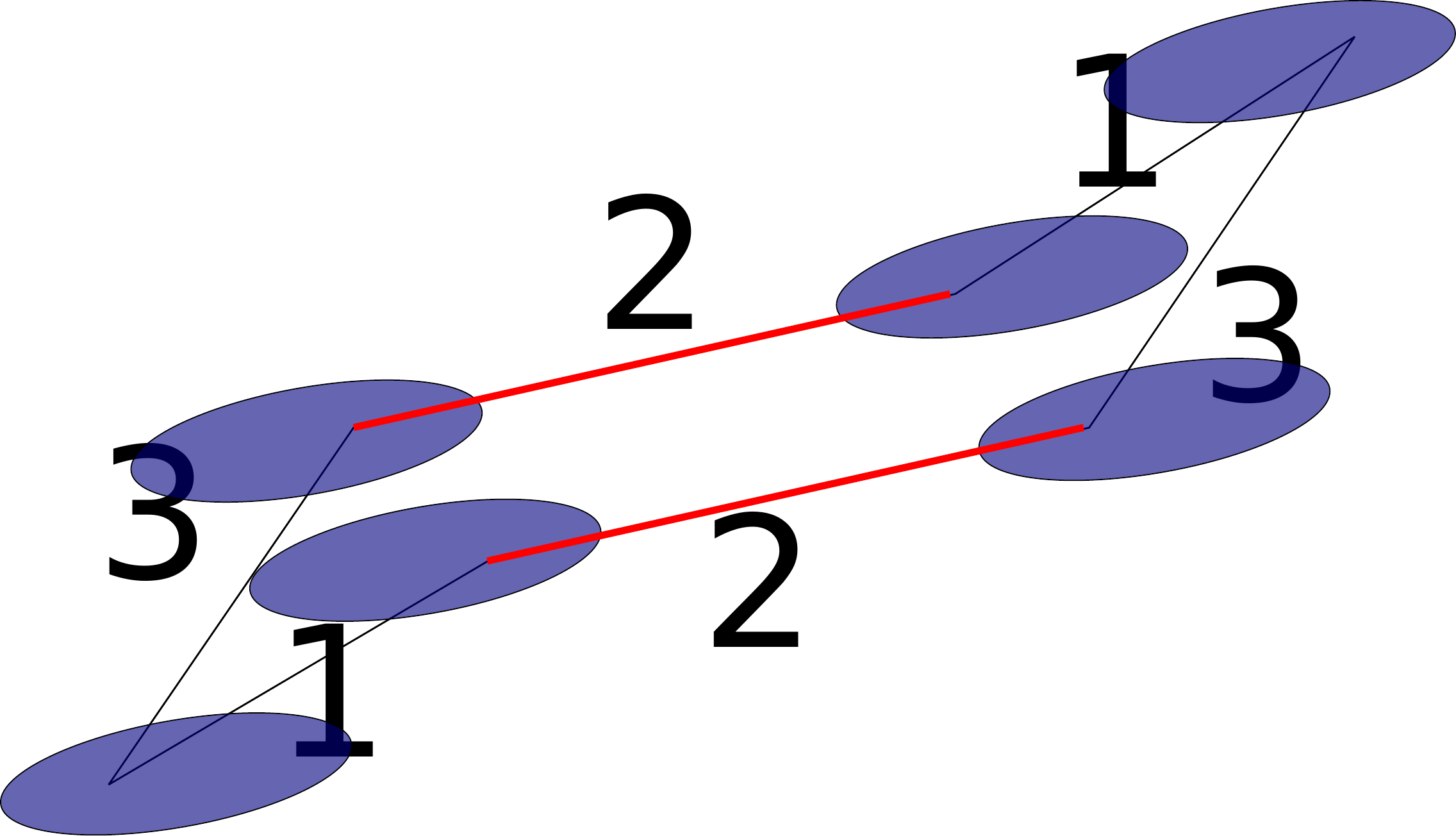}
  \caption{
    (Color Online) Illustration of the geometrical frustration introduced in this model is shown.
    The first closed loop consists of 6 variables
    and 6 variables ferromagnetically orders to ``2'' state.
    In this situation, the energy loss in the two number ``2'' bonds.
    there are no energy loss when we assign ``4'' state to this loop
    but there are direction of the loop that ``4'' state have energy loss.
  }
  \label{fig:illust-frustration}
\end{figure}

When $J_{2} = 0$,
this Hamiltonian describes
the classical ferromagnetic 4-state Potts model on the diamond lattice.
The second term gives a peculiar interaction and 
the origin of frustration
as shown in Fig.~\ref{fig:illust-frustration}.
If we consider a ferromagnetic state, \textit{i.e.}, a state
where all the $\text{K}$ atoms are in the same wavefunction $\ket{\alpha}$,
there must be always several bonds
where the two wavefunctions at the both ends are
parallel to the bond direction (Number 2 bond in Fig.~\ref{fig:illust-frustration}).
This gives a frustration.
Note that the term ``the frustration in the pyrochlore lattice''
usually indicates the electric one which comes from the tetragonal-structure network
formed by osmium and oxygen atoms.
However, the origin of the frustration in $\bpyro{K}{Os}$ described here
is completely different.
The frustration comes intrinsically from the
inter-site coupling between $\text{K}$ atoms in the diamond lattice.

\section{Method}
\label{sec:method}

The three dimensional 4-state ferromagnetic Potts model
undergoes the first order transition~\cite{Wu1982Potts}.
In order to handle the possible first order phase transition
within the Monte Carlo simulation framework,
we choose the Wang-Landau algorithm\cite{Wang2001Determining,Wang2001Efficient,Landau2002Determining,Landau2004new}.
This method enables us to calculate directly
the density of states (DOS), $g(E)$,
and also allows us to efficiently sample the ground state.
This algorithm is very effective for studying
first-order phase transitions.
The various thermodynamical quantities
are also obtained very accurately even near the first-order phase transition temperature.
Furthermore, this method can give 
estimates of the accurate ground state structure, the ground state energy and 
residual entropy, all of which are not obtained before.

This algorithm works as follows. 
Since the density of states, $g(E)$,
is \emph{a priori} not known at the very beginning of
the simulation,
we first simply set $g(E) = 1$ for all possible energies $E$.
Then we continue to update $g(E)$ until
it converges to a reasonable functional form
and the energy histogram $h(E)$ becomes flat. 
We construct a Markov chain of microscopic configuration $\{\mu\}$
using the local update scheme.
We accept  the new configuration
using the transition probability
\begin{align}
  \label{eq:35}
  p(\mu_{1} \to \mu_{2}) = \min \left[
    \frac{g(E_{\mu_{1}})}{g(E_{\mu_{2}})}, 1
  \right],
\end{align}
where $E_{\mu_{1}}$ and $E_{\mu_{1}}$ are the energy
of the system in a specific configuration $\mu_{1}$ and $\mu_{2}$, respectively.
The calculated density of states $g(E)$
and the energy histogram $h(E)$
are updated regardless of 
the acceptance of the new configuration as
\begin{align}
  \label{eq:36}
  \ln g(E_{\mu}) &\to \ln g(E_{\mu}) + \ln f_{i} \\
  h(E_{\mu}) &\to h(E_{\mu}) + 1,
\end{align}
where $f_{i}$ is a modification factor of the $i$th step of the
Wang-Landau algorithm, as defined below. At first, $f_1$ is 
chosen as $f_{1} = e$.
After some iteration,
we then check the ``flatness'' of the obtained energy histogram $h(E)$
by using a criteria that the minimum value
of $h(E)$ is not less than $80\percent$
of the average of the histogram, \textit{i.e.},
\begin{align}
  \label{eq:37}
  h(E) |_{\min}  \geq 0.8 \times \average{h(E)}.
\end{align}
When the ``flatness'' of the histogram is reached, go to the
$(i+1)$th step of the Wang-Landau algorithm,
by changing the modification factor as
\begin{align}
  \label{eq:38}
  \ln f_{i+1} = a \ln f_{i}, \\
  (0 < a < 1 ) \notag
\end{align}
where we choose $a = 0.5$, and also reset the energy histogram.
We repeat until the $i = 16$ Wang-Landau step.
The choice of $i = 16$ and $a = 0.5$
gives a reasonable convergence of $g(E)$ for the 4-state Potts model~\cite{Wang2001Determining}.

We carry out the Monte Carlo simulations with different seeds
for pseudo random number generator and calculated
the average and variance of each quantity for $J_{2} = \infty$
and $J_{2} = 10$.
We use $8^{3} = 512$ unit cells, each consists of 8 atoms
and therefore the total number is $4096$.

\section{Results}
\label{sec:results}

\begin{figure}
  \centering
  \includegraphics[width=\columnwidth,clip,keepaspectratio]{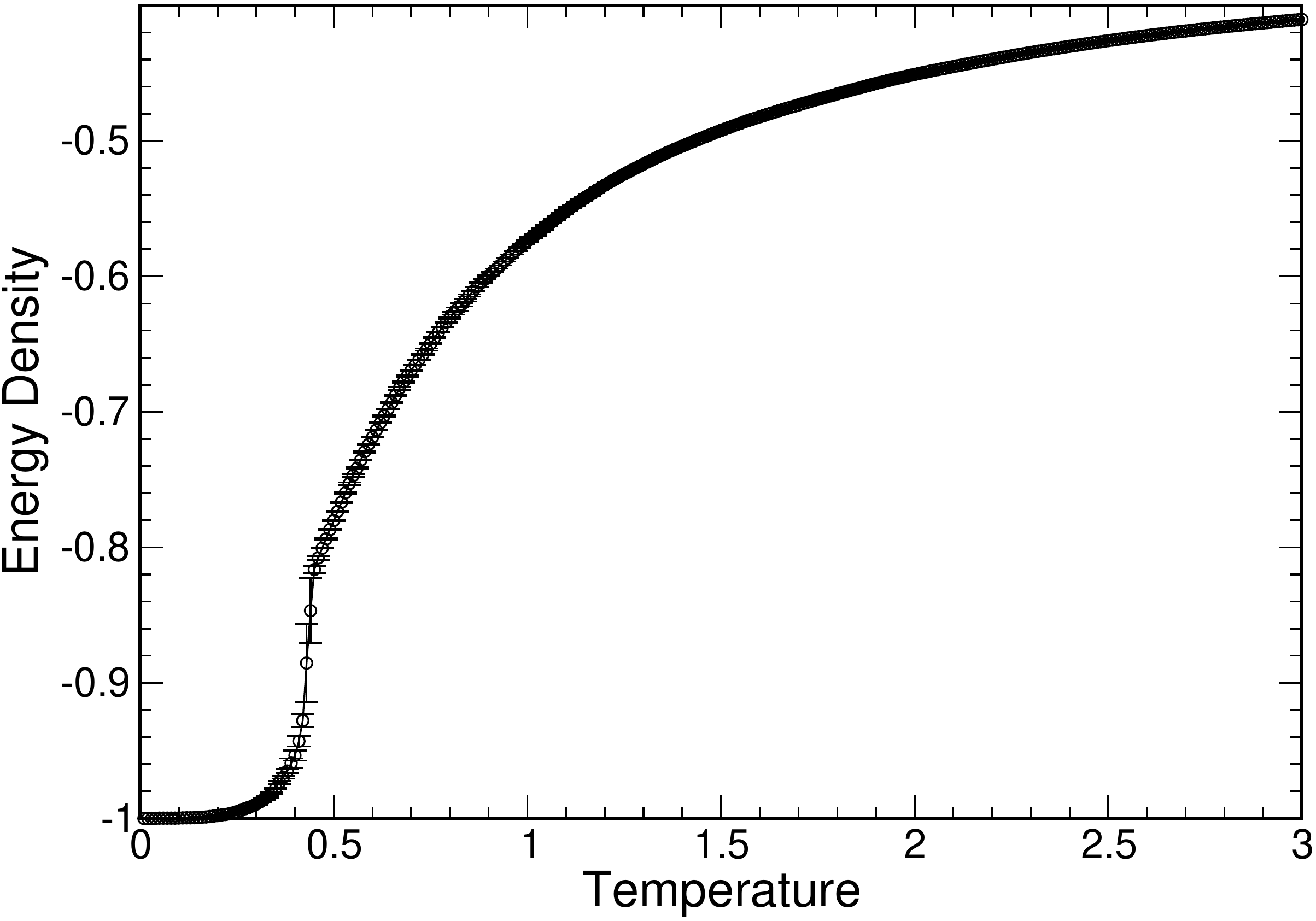}
  \caption[Energy density]{
    Energy density calculation of the extended Potts model with $J_{2} = \infty$.
    This results is of $8^{3}$ unit cells and consists of $4096$ sites.}
  \label{fig:4096ener}
\end{figure}

Figure~\ref{fig:4096ener} shows the results of energy density
for the case of $J_{2} = \infty$.
We can see a clear kink at $\temperature{} \sim 0.44$
which indicates a first-order phase transition
of the model.
We also find
that the ground state energy density is equal to $-1$ per site
within the statistical error.
If $J_{2} = 0$,
the ground state energy density
should be exactly $-2$,
because the model is reduced
to the ordinary 4-state ferromagnetic Potts model
with $2N$ bonds in the diamond lattice.
This result indicates that the ground state of the present model is
different from the simple ferromagnetic ground state 
due to the frustration induced by $J_{2}$ term.

\begin{figure}
  \centering
  \includegraphics[width=\columnwidth,clip,keepaspectratio]{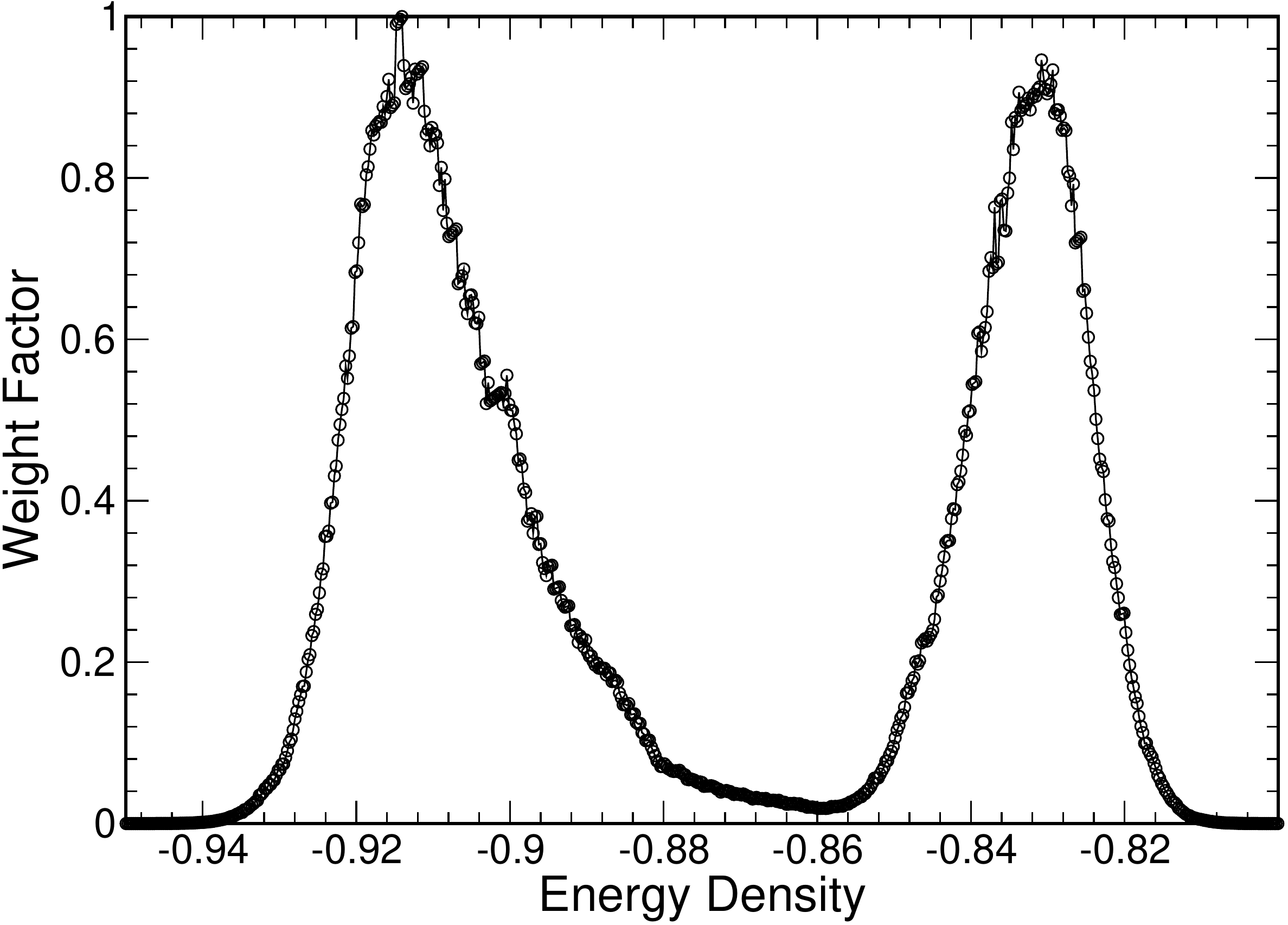}
  \caption{
    Weight factor $g(E) \exp (-\beta E)$ as a function of the energy density $E$,
    in the case of $J_{2} = 10$ near the transition temperature.
    This results is of $8^{3}$ unit cells and consists of $4096$ sites.}
  \label{fig:4096weight-finite}
\end{figure}

In order to confirm the first-order phase transition,
we study the  weight factor, $g(E) \exp (-\beta E)$,
as a function of the energy density $E$.
Figure~\ref{fig:4096weight-finite} shows
the obtained weight factor of $J_{2} = 10$ and $T = 0.435$ case.
The double-peak structure
is a clear evidence of the first-order phase transition.
We confirmed that
this double-peak structure disappears
when the temperature is only slightly off
the transition temperature, for example, at $T = 0.45$ or $T = 0.42$.

\begin{figure}
  \begin{center}
    \includegraphics[width=\columnwidth,clip,keepaspectratio]{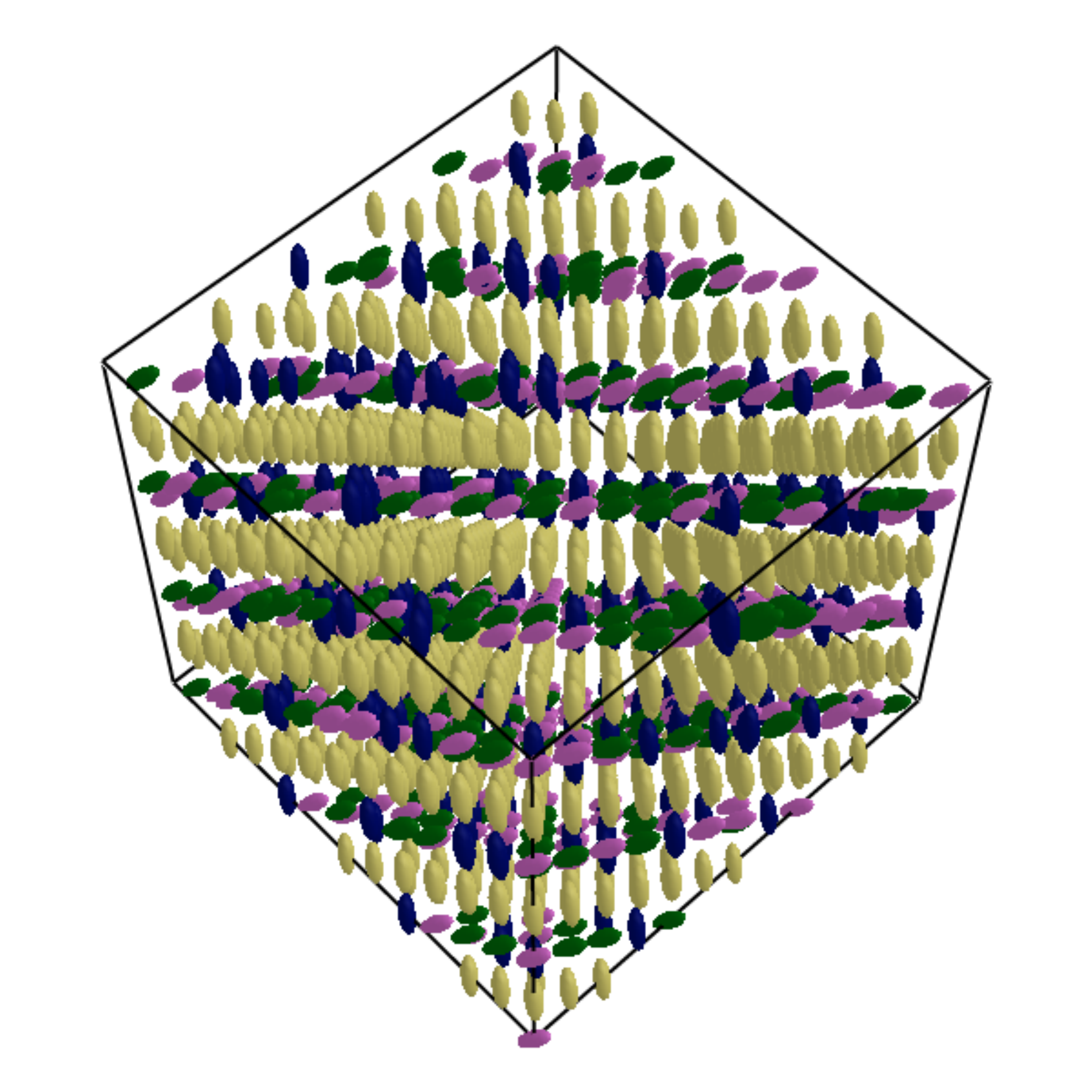}
  \end{center}
  \caption{(Color Online) Snapshot of the ground state of 6x6x6 lattice (1728 atoms). The pole direction of the ellipsoid indicate the orientation of wavefunction represented by a Potts variable..}
  \label{fig:snapshot6x6x6}
\end{figure}

A snapshot of the ground state is shown in Fig.~\ref{fig:snapshot6x6x6}.
We can see that half of the wavefunctions $\ket{\alpha}$ represented by Potts variables
order in a two-dimensional hexagonal sheet-like structure,
in which half sites are located slightly above the sheet
and the remaining half slightly below.
This sheet-like plane is perpendicular
to the majority rattling direction.
The sandwiched sheets between the directions of the majority wavefunction remain disordered
consisting of the other three wavefunctions.
This ground state structure is a very new and peculiar ground state structure
induced by an interesting frustration interaction in the present model.

\begin{figure}
  \begin{center}
    \includegraphics[width=\columnwidth,clip,keepaspectratio]{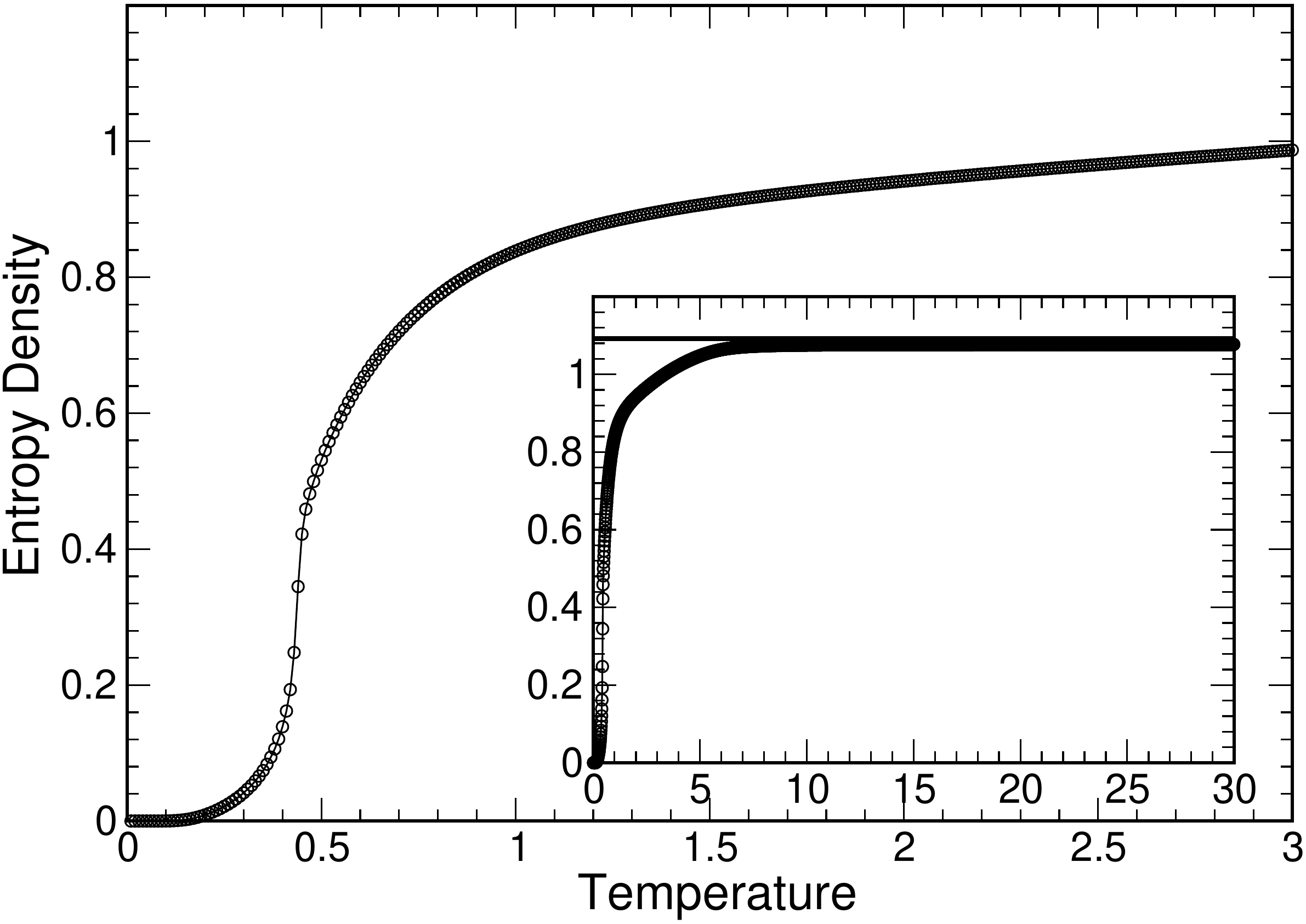}
  \end{center}
  \caption[Entropy Density]{
    Entropy density calculations of the extended Potts model with $J_{2} = 10$
    are shown.
    These results are of $8^{3}$ unit cells and consists of $4096$ sites
    and the calculation up to the temperature $\temperature{} = 3$ is shown.
    The jump of the Entropy density is observed at $\temperature{} \sim 0.44$
    which corresponds to the first order phase transition temperature.
    The inset shows the calculation up to the temperature $\temperature{} = 30$.
    The line denotes the value of $1.092$.
  }
   \label{fig:4096entropy-finite}
\end{figure}

\begin{figure}
  \centering
  \includegraphics[width=0.70\columnwidth,clip,keepaspectratio]{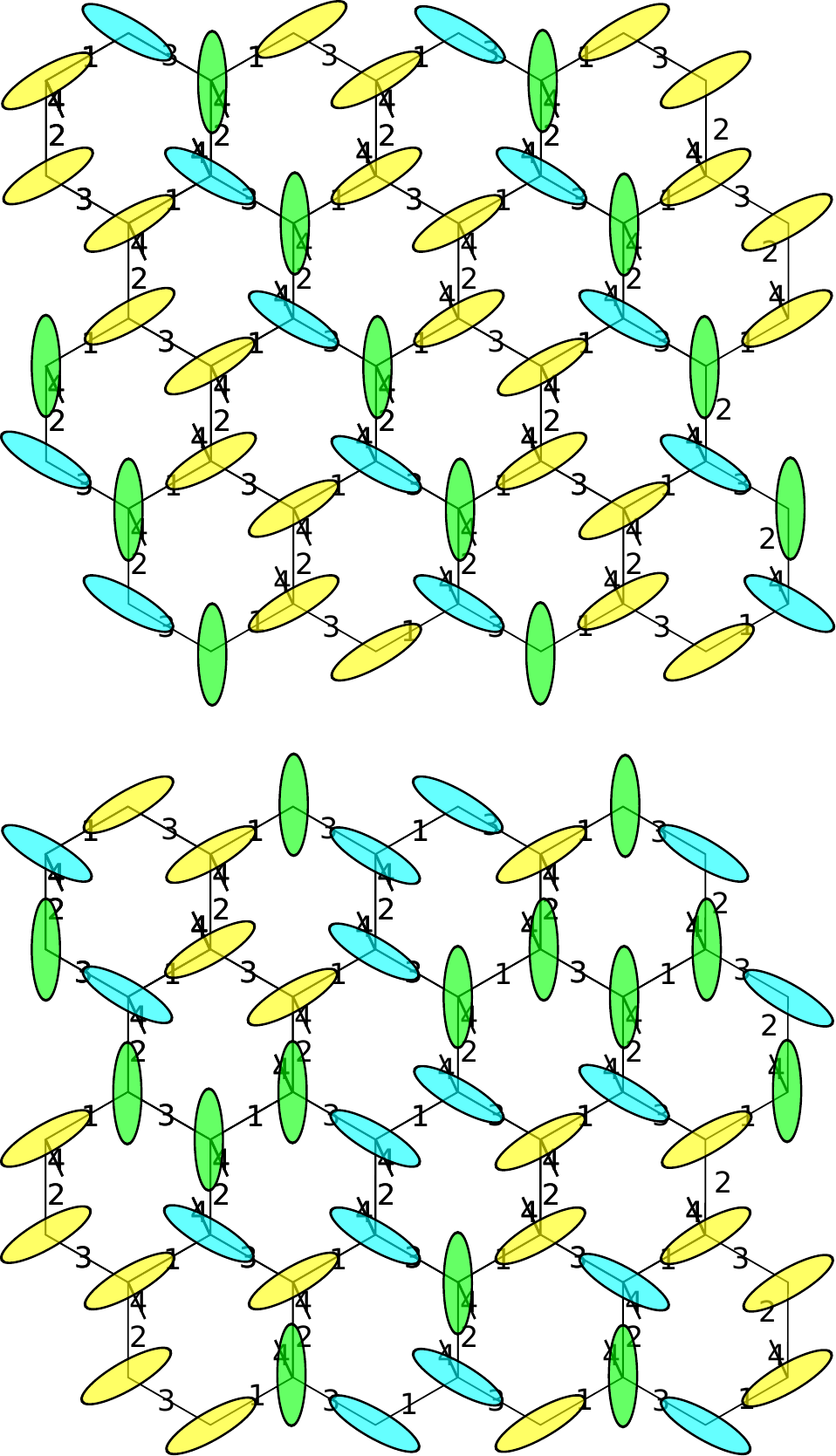}
  \caption{(Color Online) Two examples
    of the ground state configurations in a disordered sheet are shown.
    Both configurations satisfy periodic boundary condition.
    The yellow, green and blue ellipse denote
    Potts variable 1, 2 and 3, respectively.}
  \label{fig:ferroplane}
\end{figure}

Let us now discuss the residual entropy density originating
from the disordered sheets.
The obtained entropy density as a function of $\temperature{}$
in the case with $J_{2} = 10$ is shown in Fig.~\ref{fig:4096entropy-finite}.
We find that the entropy density in the $\temperature{} \to \infty$ limit
converges to $1.092(6)$ when we set the ground-state entropy to be zero.
The correct entropy density in the $\temperature{} \to \infty$ limit should become
\begin{align}
  \label{eq:16}
  \average{S} (\temperature{} \to \infty) = \log 4,
\end{align}
since every site has 4 degrees of freedom in the present model.
Therefore the residual entropy density in the present model is calculated as
\begin{align}
  \label{eq:24}
  \average{S_{0}} = \log 4 - 1.092 =  0.294.
\end{align}
This residual entropy density must come from the degenerate ground state
of the disordered sheets as shown in Fig.~\ref{fig:ferroplane}.

In the following, we consider the residual entropy in detail.
Two examples of the ground state configurations are shown in
Fig.~\ref{fig:ferroplane}.
Here the bonds with ``bond number'' 4 are perpendicular to the hexagonal sheet,
and it is assumed that all the sites in the nearest-neighbor sheet have
Potts variable 4.
Therefore, the Potts variable 4 is forbidden in the sheets shown in Fig.~\ref{fig:ferroplane}.
Therefore, if the Potts variables in the disordered sheets with $N / 2$ sites
are completely random,
the total number of possible configurations is $3^{\frac{N}{2}}$ and the
corresponding entropy density is
\begin{align}
  \label{eq:6}
  \average{S'_{0}} = \frac{1}{N} \log 3^{\frac{N}{2}} = 0.5493.
\end{align}
However, the residual entropy density
in eq.~\eqref{eq:24}
is much smaller than that calculated in eq.~\eqref{eq:6}.
The origin of this difference
comes from the constraint inside the disordered sheet.

Actually,
a bond direction and the two states at the both ends of the bond
cannot be all parallel inside the disordered sheet.
Even under the constraint,
there are many possible configurations in the ground state.
The upper figure and the lower figure of Fig.~\ref{fig:ferroplane}
look very different
but they have exactly the same energy.
The upper figure has a site-number unbalance. The $\frac{N}{4}$ sites are
filled with Potts variable 1
whereas only $\frac{N}{8}$ sites are filled with Potts variable 2 and 3.
In the lower figure, the sites are almost equally filled with the Potts variables
1, 2 and 3.
Note that the number of the spins in the sheet shown in Fig.~\ref{fig:ferroplane}
is not dividable by 3 but there is no difficulty of filling Potts variables almost equally.

The effect of the constraint inside the sheets
can be taken into account approximately as follows~\cite{Udagawa2002Exact}:
Let us start from the completely random configurations
which consists of $3^{\frac{N}{2}}$ states.
If we focus on a single bond in the hexagonal sheet,
we notice that the forbidden states for that bond are included.
Since the configurations are chosen completely randomly,
the probability of appearance of this forbidden state is
$\frac{1}{9}$.
Here $3 \times 3 = 9$ represents the total number
of the states at the both ends of the
corresponding bond.
Therefore, for each bond, $\frac{1}{9}$ configurations
should be discarded.
As a result, the total number of the allowed configurations
can be estimated as
\begin{align}
  \label{eq:26}
  3^{\frac{N}{2}}
  \left(
    1 - \frac{1}{9}
  \right)^{\frac{3}{4}N}
  =
  \frac{8^{\frac{3}{4}N}}{3^{N}},
\end{align}
where $\frac{3}{4}N$ is the number of bonds
in the sheet.
The resulting entropy density is given by
\begin{align}
  \label{eq:29}
  S_{0} = \log
  \left(
    \frac{8^{\frac{3}{4}}}{3}
  \right)
  = \log (1.586) = 0.461,
\end{align}
which is closer to the numerically obtained residual entropy density,
$0.294$, than the completely random value $0.5493$.

\section{Discussion and Conclusions}

Let us compare the present results with experiments of 
$\bpyro{K}{Os}$.
Despite the various peculiar properties of the present model,
it apparently shows directional symmetry breaking, which 
is not observed in $\bpyro{K}{Os}$.
The transition temperature of the model,
$0.44 J_{1} \sim 80\kelvin$,
is of order high
compared with the experimentally obtained
first order transition temperature of $\temperature{p} = 7.5\kelvin$.
Thus, the simplified model proposed by Kun{\v e}s \textit{et al.}
does not explain
the rattling transition in $\bpyro{K}{Os}$.
A more sophisticated model Hamiltonian will be necessary.
Recently, Hattori and Tsunetsugu~\cite{Hattori2009Possible}
proposed another model for this rattling transition of $\bpyro{K}{Os}$.
They argued the rattling transition
by introducing a fifth state
in addition to the four states discussed above.
Although they succeeded to explain the first-order phase transition
without symmetry breaking,
the physical origin of the fifth state is not clear.
Moreover, a recent experiments shows
that the lattice expands below $T_p$,
which contradicts their prediction.
Thus the rattling transition in $\bpyro{K}{Os}$ remains an open question.

Although the present model is not applied to the transition in $\bpyro{K}{Os}$,
we found that this model has very peculiar features.
The obtained phase transition is of
first order, which is verified
from the double peak structure of the weight factor shown in Fig.~\ref{fig:4096weight-finite}.
The ground state spin configuration of this model shows that
half of the spins in the system are ordered
and form a hexagonal-sheet-like structure.
The overall possible structures of the ground state snapshot configuration are illustrated
in Fig.~\ref{fig:snapshot6x6x6} and Fig.~\ref{fig:ferroplane}.
The remaining half of the spins are distributed disordered
even in the low temperature limit $T \to 0$,
which gives the residual entropy density of $0.294(6)$.
These results indicate a new kind of interesting ground state which 
will be worth further studying as a result of new type of frustration.
We use the ALPS library for calculation~\cite{Albuquerque2007ALPS}.

%% JPSJ style
%\section*{Acknowledgment}
%% REVTeX style
\begin{acknowledgments}
We are grateful to Dr. Todo (Department of Applied Physics, University of Tokyo)
for allowing to use the part of the ALPS parapack library.
This work is supported in part by Global COE Program ``the Physical Sciences Frontier'', MEXT, Japan.
The computation in this work has in part been done using the facilities of the Supercomputer Center,
Institute for Solid State Physics, University of Tokyo
and the Next Generation Super Computing Project, Nanoscience Program, MEXT, Japan.
%% REVTeX style
\end{acknowledgments}

%% JPSJ style
%\bibliographystyle{_jpsj}
%% REVTeX style
\bibliographystyle{apsrev}
\bibliography{rigarash}

%% JPSJ style
%\label{lastpage}
\end{document}